\documentclass[aps,ajp,showpacs,twocolumn,superscriptaddress]{revtex4-2}
\usepackage{epstopdf}
\usepackage{graphicx}
\usepackage{dcolumn}
\usepackage{bm}

\usepackage[pdfpagemode=UseNone,pdfstartview=FitH,colorlinks=true,linkcolor=blue,urlcolor=blue,anchorcolor=blue,citecolor=blue]{hyperref}
\usepackage{cancel,xcolor}
\usepackage{soul,xcolor}
\setstcolor{red}

\usepackage[utf8]{inputenc}
\usepackage[T1]{fontenc}
\usepackage{mathptmx}
\usepackage{etoolbox}

\usepackage{comment}
\usepackage{braket}
\usepackage{physics}
\usepackage[normalem]{ulem}

\usepackage{color}

\begin{document}

\title{ Wannier-Stark localization, confinement and edge states}


\author{N. Aucar Boidi}
\email{nairaucar@gmail.com}
\affiliation{The Abdus Salam International Centre for Theoretical Physics, Strada Costiera 11, I-34151, Trieste, Italy}

\author{A. Aharony}
\email{aaharonyaa@gmail.com}
\affiliation{School of Physics and Astronomy, Tel Aviv University, Tel Aviv 6997801, Israel}
\email{aaharonyaa@gmail.com}

\author{O. Entin-Wohlman}
\email{orawohlman@gmail.com}
\affiliation{School of Physics and Astronomy, Tel Aviv University, Tel Aviv 6997801, Israel}
\email{orawohlman@gmail.com}

\author{C. R. Proetto}
\email{crproetto@googlemail.com}
\affiliation{Centro At\'{o}mico Bariloche, Instituto Balseiro, Instituto de Nanociencia y Nanotecnolog\'{\i}a CNEA-CONICET, Bustillo 9500, 8400 Bariloche, Argentina}


\vspace{1cm}

\date{\today}

\begin{abstract} 

 The boundary between a system's bulk and the vacuum can be modeled by  a potential which confines the electrons to the bulk. Here  we present the example of the Wannier-Stark linear potential, generated by an electric field along one axis of a two-dimensional lattice. Along that axis, the potential generates electronic states which are localized around each lattice site, with eigenenergies which form the Wannier-Stark ladder. In the transverse direction, the states are governed by a tight binding model with free Bloch states. In the ground state, filling these states under the Pauli principle generates an insulating bulk and an edge which can be metallic in the transverse direction. This  paper explains the concepts of localization, localization length, filling and confinement.  The paper also acquaints the readers with the use of Bessel functions in the solution of a simple physical model. 
 These topics can be easily included in courses on solid state physics, and only require prior knowledge of quantum mechanics.

\end{abstract}

\maketitle

\section{Introduction}

 Many recent studies~\cite{r1,r2,r3} concern the interface between the bulk of a solid and the external empty vacuum. In addition to the quantum states occupied by electrons in the bulk of such systems, there often arise additional ``edge" states on the boundaries. Much of the recent interest concern topological edge states arising due to strong magnetic fields in two-dimensional quantum Hall systems~\cite{top,BIH}. In these systems, the electronic states in the bulk are localized, so that the bulk is an insulator, and the edge states on the boundary are extended, allowing the flow of current along the boundary. These concepts, ``localized" and ``extended", will be explained in detail below.  
 
 In the present paper, we consider a non-topological situation, without a strong magnetic field. Even without the topological aspects, the boundary can exhibit a variety of interesting edge states~\cite{Khanna2022}. Such states can be tested experimentally, e.g., in arrays of cold atoms 
\cite{Wilkinson1996,Guo2021}. 
We present here a simple analysis of non-interacting electrons which move on a periodic lattice and are confined by a potential. 
Electron-electron interactions are 
briefly mentioned at the end of the paper. Our simple model, which can be solved analytically, yields localized states in the bulk and extended states along the two-dimensional boundary.

Some theoretical studies describe such edges by a confining harmonic potential~\cite{Heidrich2010}, with wave functions which decay as they enter the vacuum region (where their energy is below this potential). Here we consider an alternative model: the electrons feel an electric field perpendicular to the edge,  generating a linear potential, termed  the  ``Wannier-Stark potential"~\cite{basics,Wannier59,fukuyama}. When this potential grows from left to right, the electrons are confined to the left side of the sample, which represents the ``bulk". For simplicity, our analysis begins with a one-dimensional chain of (discrete) sites 
(Fig.~\ref{111} below). The extension to two dimensions follows, together with the introduction and discussion of edge states.

In one dimension, the eigenfunctions of the Wannier-Stark potential are known to be the Bessel functions of the first kind~\cite{fukuyama,Bessel,Arfken2012,Bowman} and the eigenenergies form a ladder, with a constant gap between consecutive eigenenergies.  For a single-particle state, as shown in Fig.~\ref{Fig1}, these functions are localized around each site of the chain, introducing  the important concept of ``localization".  The analysis below  uses these important special functions to calculate the related localization lengths  (also shown in Fig.~\ref{Fig1}). 

 After introducing the Wannier-Stark states for the infinite one-dimensional chain, we consider a finite chain, and the filling of these states by electrons in the ground state, including the spin degrees of freedom and imposing the Pauli principle (which allows at most two electron in each electronic state). In one dimension, the resulting occupation profile is shown in Fig. \ref{Occupation}: far away from the boundary into to bulk (on the left side) the profile is flat, with two electrons per site. The profile decays gradually to zero into the vacuum (on the right side), with a rate which is associated with the localization length of the Wannier-Stark states.

 We then study a two-dimensional lattice, formed by stacking the above linear chains and adding tunneling between them. Each eigenenergy on the Wannier-Stark ladder is now replaced by a band of energies associated with the extended electronic states in the transverse direction, see Fig. \ref{Fig3}. Filling these states up to the Fermi energy implies that only the ``last" electrons, filling the rightmost ``edge" states, can ``move" along the bulk-vacuum boundary.

We begin with a brief review of the tight-binding approximation in Sec. \ref{II}, including the example of the energy bands of electrons under a periodic potential, called Bloch electrons (Sec. \ref{Bloch}). The linear Stark potential is then added in Sec. \ref{III}.
The localization of the resulting eigenfunctions, related to the Bessel functions, is then discussed in Sec. \ref{loc}. In particular, we introduce  the ``localization length", which is an important quantity in modern condensed-matter physics. The many-electron case, utilizing the Pauli principle for the  filling of the lattice, is then described in Sec. \ref{MB}.  Finally, Sec. \ref{ES} describes the two-dimensional edge states,  and Sec. \ref{dis} gives a brief qualitative review of the effects of electron-electron interactions. The Appendix contains a few exercises.


\section{The Tight-binding Approximation}\label{II}

One of the simplest models to describe non-interacting electrons on a lattice is the tight-binding approximation~\cite{ashcroft-mermin,Powell,book}. Below we mention its key ingredients, as needed for the understanding of the following sections. 
One starts with  single atoms located at the $N$ sites of a periodic lattice.  The simplest configuration consists of one relevant orbital, with energy $\varepsilon^{}_0$ and with eigenstate $|j\rangle$ at site $j$, which can be represented by an $N$-component vector with component $1$ at row $j$ and zero otherwise. 
 The off-diagonal Hamiltonian matrix elements ${\cal H}^{}_{mj}={\cal H}^{}_{jm}=\langle j|{\cal H}|m\rangle$ correspond to the overlap between wave functions at sites $j$ and $m$ due to the potential of ions on the lattice.
The Schr\"odinger equation, ${\cal H}|\Psi\rangle={\cal E}|\Psi\rangle$, for the $N$ coefficients of $|\Psi\rangle=\sum_{j=1}^N \psi(j)|j\rangle$ can be written as
\begin{equation}
{\cal H}^{}_{jj}\psi(j) + \sum_{m\ne j}{\cal H}^{}_{jm}\psi(m) = {\cal E}\psi(j)\ ,
\label{GenH}
\end{equation}%
where  the diagonal matrix element ${\cal H}^{}_{jj} $ may contain a local potential energy. 
For simplicity, we set 
the tunneling energy ${\cal H}^{}_{jm}\equiv -t$  when $m$ is a nearest-neighbor of $j$ (and zero otherwise) 
and consider  
a one-dimensional lattice, with $N=2M+1$ and $j=-M,~-M+1,~\ldots,~M$. Finally,  Eq.~(\ref{GenH}) becomes
\begin{equation}
{\cal H}^{}_{jj}\psi^{}_\ell(j)-t\psi^{}_\ell(j-1)-t\psi^{}_\ell(j+1) = {\cal E}^{}_\ell\psi^{}_\ell(j).
\label{TB}
\end{equation}
Here, the index $\ell $ enumerates the eigenstates; the finite chain must have $N$ discrete egienenergies, 
which may also depend on the  boundary conditions. The chain has a length $L=2Ma$, with $a$ being the lattice constant.


\vspace{-4mm}

\subsection{ Bloch electrons} \label{Bloch}

Without a potential energy, we set ${\cal H}^{}_{jj}=0$. Apart from the boundaries,  the eigenstates of Eq. (\ref{TB}) are the  (orthonormal) ``plane waves" $\psi^{}_{\ell}(j)=\exp[ik^{}_{\ell}aj]/\sqrt{N}$, with the eigenvalues 
${\cal E}^{}_\ell=-2 t \cos(k^{}_\ell a)$, and the possible values of $k^{}_\ell$ are determined by the boundary conditions. 
 For example, periodic boundary conditions for a chain of $N$ sites require $\psi^{}_{\ell}(j)=\psi^{}_{\ell}(j+N)$, and then $e^{ik^{}_{\ell}aj}=e^{ik^{}_{\ell}a(j+N)}$, which is obeyed for $k^{}_\ell=2\pi\ell/(Na)$, $\ell=-M,~-M+1,~\ldots,~M$. 
 The $N-$fold degenerate eigenenergy ${\cal E}=\varepsilon^{}_0=0$ is now replaced by a {\it band} of 
energies in the range $-2t\le {\cal E}^{}_{\ell}< 2t$, which become very dense for very large $N$. Note: for $k^{}_\ell\ne 0$ we still have a degeneracy between the right-moving and left-moving waves, $\pm |k^{}_\ell|$. The solutions obey $\psi^{}_\ell(j)=\psi^{}_{\ell+N}(j)$, and therefore it is sufficient to study them in 
the first Brillouin zone, $-\pi/a<k^{}_\ell<\pi/a$. 




\vspace{-4mm}

\subsection{The Wannier-Stark ladder}\label{III}

\begin{figure}
\centering
\includegraphics[width=0.4\textwidth=1]{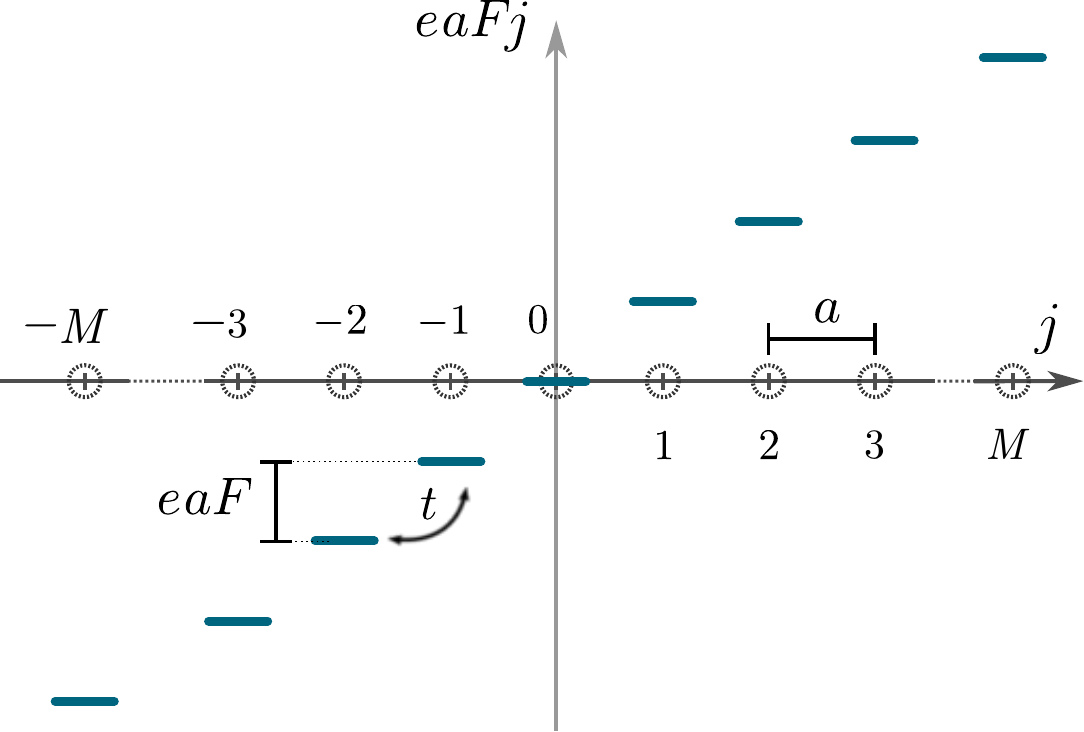}
\caption{ A sketch of the one-dimensional lattice, represented by the dotted circles along the horizontal axis, with an odd number of sites placed in a linear potential $\mu^{}_j=eaFj$. 
Each step indicates the energy of the corresponding site due to the presence of the potential which vanishes at the central site of the chain $j=0$. 
Electrons  can hop between neighboring sites with energy $t$.} 
\label{111}
\end{figure}

An electric field $F$ along the direction of the one-dimensional lattice adds to the Hamiltonian a potential-energy term, ${\cal H}^{}_{jj}=eaF j$,  acting on the electron at the site $j$ (See Fig. \ref{111}).  Here $-e$ is the charge of the electron.  This potential breaks the periodicity of the problem, and thus competes with the periodic potential discussed above.  Clearly, periodic boundary conditions are no longer possible. Since we consider the states in the middle of an infinite lattice,  the boundary conditions do not matter.
Equation (\ref{TB}) is  now replaced by
 \begin{align}
{\cal E}^{}_{\ell}\psi^{}_{\ell}(j)
=eaFj\psi^{}_{\ell}(j)-t\psi^{}_{\ell}(j-1)-t\psi^{}_{\ell}(j+1)\ .
\label{TB2}
\end{align}
In the absence of the tunneling term,  $t=0$, the eigenstates are fully localized on the sites (i.e., one state per each site), and the wave function $\psi^{}_\ell(j)=\delta^{}_{j\ell}$ has the eigenvalue ${\cal E}_\ell=eaF\ell$, with integer $\ell$. We now show that this ladder of equidistant levels, called the ``Wannier-Stark ladder", persists also for $t\ne 0$.

We start with a specific eigenstate of the Wannier-Stark problem, $\psi^{}_\ell(j)$, with eigenenergy ${\cal E}^{}_\ell$, which obeys Eq. (\ref{TB2}). In the limit $N\rightarrow \infty$, where boundary conditions  are not necessary, we can shift the whole lattice by one unit, and define a new wave function, $\overline{\psi}^{}_\ell(j)=\psi^{}_\ell(j-1)$. It is straightforward to see that $\overline{\psi}^{}_\ell(j)$ satisfies the same Eq.~(\ref{TB2}), but with a new energy, $\overline{\cal E}^{}_\ell={\cal E}^{}_\ell+eaF$. Repeating this procedure shows that the energy levels form a ladder with the energy step $eaF$, ${\cal E}^{}_\ell=eaF\ell$, and that the energy of $\overline{\psi}^{}_\ell(j)$ is the next step on that ladder, i.e.,  $\overline{\cal E}^{}_\ell={\cal E}^{}_{\ell+1}$ \cite{fukuyama}. 
This also proves that 
$\psi^{}_{\ell}(j-1)=\overline\psi^{}_\ell(j)=\psi^{}_{\ell+1}(j)$.  Repeating this equality and reducing $j$ and $\ell$ gradually, this becomes $\psi^{}_\ell(j)=\psi^{}_{\ell-j}(0)$. This proves that the eigenstates $\psi^{}_\ell(j)$ depend only on the difference $\ell-j$. 
Unlike the Bloch electron, whose energies form a continuum  band in the limit $N\rightarrow\infty$, the energies of the Wannier-Stark ladder have finite gaps between them.

For convenience, we introduce  dimensionless quantities.
Dividing Eq. (\ref{TB2})  by $t$, and using ${\cal E}^{}_\ell(x)/t=2\ell/x$, with
\begin{align}
x=2t/(eaF)\ ,
\end{align}
one finds
\begin{align} 
\psi^{}_\ell(j-1)+\psi^{}_\ell(j+1)=[2(\ell-j)/x]\psi^{}_\ell(j)\ .
\label{TB88}
\end{align}
In this dimensionless parametrization, both the eigenenergies ${\cal E}^{}_\ell/t$ and eigenfunctions $\psi^{}_\ell(j;x)$ depend only on the dimensionless parameter $x$.

Using the replacements $\psi^{}_\ell(j)=\psi^{}_{\ell-j}(0)$, this equation becomes
\begin{align}
\psi^{}_{\ell-j-1}(0,x) + \psi^{}_{\ell-j+1}(0,x) = [2(\ell-j)/x] \psi^{}_{\ell-j}(0,x)
\end{align}
 This equation turns out to be  identical to the Bessel functions equation~\cite{fukuyama,Bessel,Arfken2012,Bowman}
\begin{align} 
J_{\nu-1}^{}(x)+ J_{\nu+1}^{}(x)=[2\nu/x] J_{\nu}^{}(x)\  ,
\label{id}
\end{align}
where $J^{}_\nu(x)$ is the Bessel function of the first kind, of integer order $\nu=\ell-j$,
 provided we identify the eigenfunctions of the Wannier-Stark problem, $\psi^{}_\ell(j;x)$,  with the Bessel functions $J^{}_{\ell-j}(x)$,
\begin{align}
\psi^{}_\ell(j;x) =\psi^{}_{\ell-j}(0;x)\rightarrow J^{}_{\ell-j}(x)\ .
\label{psiJ}
\end{align}
This identification proves that the eigenfunctions of the Wannier-Stark potential on a periodic lattice with nearest-neighbor tunneling are indeed the Bessel functions of the first kind. For $N\rightarrow\infty$ these wave functions are orthonormal, as in Eq.  (\ref{ort1}) \cite{Bessel}. In the eigenstate $|\Psi^{}_\ell\rangle$, the probability of finding the electron at site $j$ is thus $J^{}_{\ell-j}(x)^2$.
For integer $\nu$ the Bessel functions obey  $J^{}_{-\nu}(x)=(-1)^\nu J^{}_\nu(x)$, and therefore it is sufficient to study the solutions only for 
$\ell-j\ge 0$~\cite{Bessel}.

\begin{figure}
\centering
\hspace{2cm}
\ \ \ \ \ \includegraphics[width=0.7\textwidth=1]{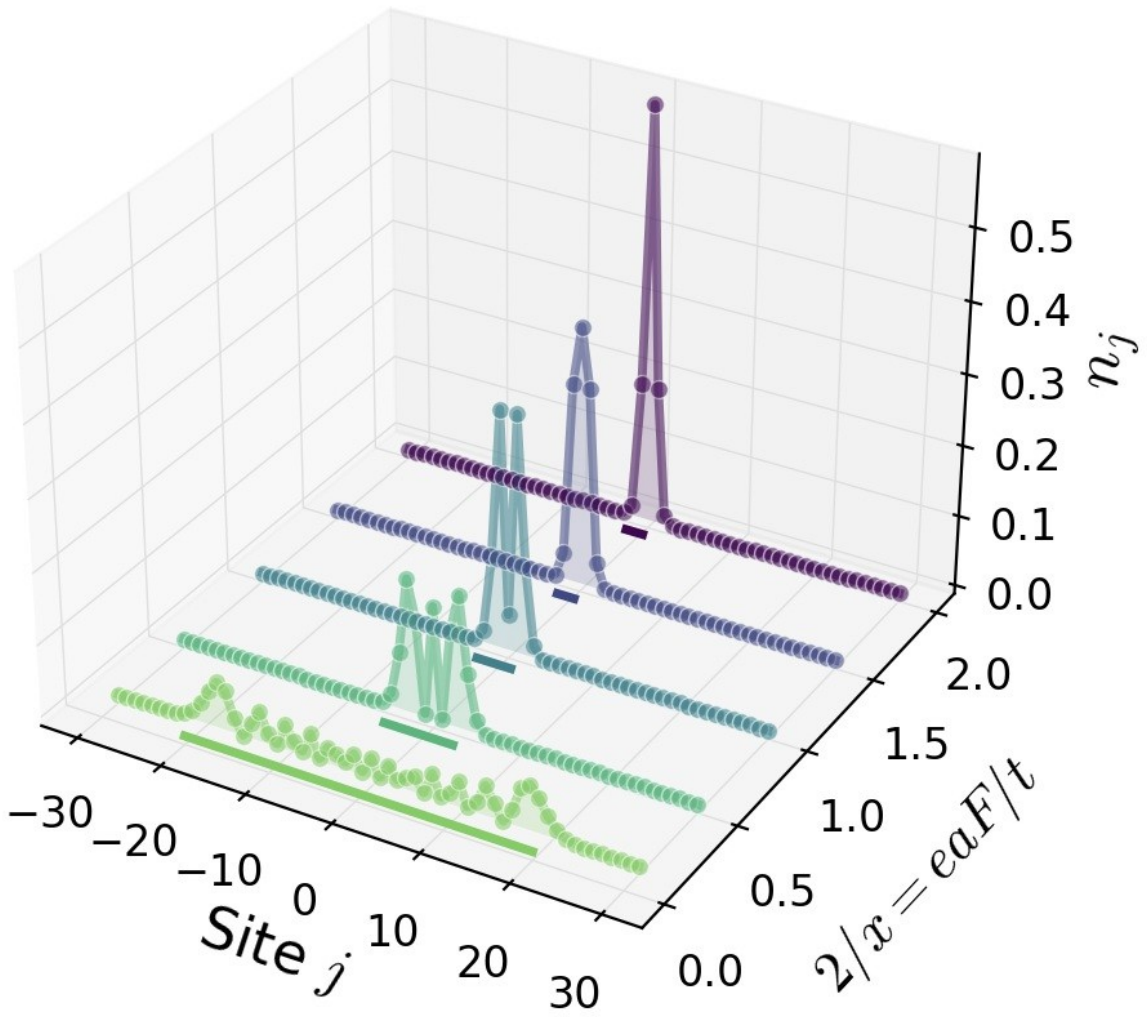}
\caption{(Color online) Probability $n^{}_j=J^{}_j(x)^2$ of finding a single electron in the Wannier-Stark eigenstate $\ell=0$, as a function of site index $j$, for several values of the electric fields $F$, in units of $t/(ea)$. These states are localized around the cite $j=0$. The full lines connect the actual values, which are calculated only for integer $j$. The straight segments signal the value of the localization length $2\xi=4t/(eF)$.} 
\label{Fig1}
\end{figure}

 For the eigenstate with $\ell=0$, the probability of finding the electron on site $j$, equal to $J^{}_j(x)^2$, is shown in Fig. \ref{Fig1} for several values of $eaF/t=2/x$. From Ref.~\cite{Wilkinson1996} we obtain that typical
values for the lattice constant, effective electric field, and hopping
parameters for the case of optical lattices are as follows: $a \sim
3000$\AA, $F \sim 4 \times 10^{-6}$ V/cm, and $t \sim 40 \times
10^{-12}$ eV. It is then seen that $x = 2t/eaF \sim 1$. However, the values of $x$ can be easily tuned by varying any of these parameters.

  In the limit of a very large electric field $F$ (i.e. small $x$), $J^{}_{\ell-j}(x)\rightarrow J^{}_{\ell-j}(0)=\delta^{}_{j,\ell}$~\cite{Bessel}, implying that the wave function $\psi^{}_\ell(j;x)$ is fully localized at the site $j=\ell$ (as in the zero tunneling case, $t=0$).
 As the electric field decreases, the electron's wave function spreads over more sites and the localization length, over which the wave function differs significantly from zero (see below), broadens. In fact, for large $j$ and $\ell=0$ the Bessel function 
  decays more strongly than exponentially in $j$  \cite{fukuyama,Bessel} 
  and the wavefunction is localized in a region around $j=0$. 
 As long as this region is far from the boundaries, the solution (which was derived for an infinite chain) remains approximately correct also in the finite samples.  In fact, a steep wall to the left of the sample, which implies $\psi_{-M-1}=0$, affects only Eq. (\ref{TB88}) for $j=-M$. As long as the localization length is smaller that $Ma$, the states near $j=0$ will not be affected.  As $F$ decreases, the internal region of the wave function becomes more and more similar the wave function of the wave function of the ``free" electron with hard walls (Exercise 2 in App. \ref{App}), oscillating with almost a constant ``envelope". Indeed, in the limit $F\rightarrow 0$ the solution should approach that wave function.


\section{The localization length}\label{loc}

A famous example of localized states concerns the electronic states in disordered crystals, with random values of onsite energies, which undergo  Anderson localization~\cite{Anderson1958}.  In one and two dimensions, any randomness causes localization of all the electronic states, but quantitative analyses require averages over many random realizations. In higher dimensions there arises a metal-insulator transition between extended and localized states ~\cite{anloc}. Another example concerns the bulk states in the quantum Hall insulator~\cite{top}.
The size of the region in which the electronic state is localized is termed  the  ``localization length". 

The Wannier-Stark ladder states are a simple example of  localization without disorder (i.e., without randomness). 
Below we present two estimates of the localization length for  the Wannier-Stark states with $\ell=0$, which are centered on site $j=0$. Both  estimates turn out to be linear in $xa$, with coefficients of order 1.  The numerical differences are of no special importance, as both demonstrate the increase of the localization length with decreasing electric field.


\vspace{-4mm}

\subsection{Rough estimate of the width}

One estimate of the localization region width follows from the argument in
Ref. \cite{coex}: the Hamiltonian contains a competition between two main energy scales, the tunneling energy $t$ and the Wannier-Stark energy $eFa\ell$. Consider now the state centered on $j=0$. When $|eFaj|\ll 2t$, the electric field term in Eq. (\ref{TB2}) is small, and the wave functions in this region may be close to those of the ``free" electron, which are extended over this region  (See Fig. \ref{Fig1}). Outside of this region the electric field becomes important, and the wave functions decay. Therefore, a rough estimate of the region within which the wave functions are extended is given by $|j|\le x=2t/(eaF)$, and the half-width of this range is roughly
\begin{align}
    \xi^{}=xa
    =2t/(eF)\ .
    \label{dw}
\end{align}
 The straight segments in Fig. (\ref{Fig1}) show $2\xi$, and the good agreement with the visible width of the wave function confirms this rough estimate.
 As expected, this width decreases as the potential becomes steeper. In fact,  dimensional analysis shows that $\xi/a$ can only depend on $x$, which is the only dimensionless combination  of the parameters in our problem. However, the linear dependence in Eq. (\ref{dw}) is not trivial. Indeed, other estimates of this width also increase  linearly with  $x$.



\vspace{-4mm}

\subsection{Mean square average}

A natural definition of the localization length of a single-particle wave function is given by the mean square average of the distance from the center of the localized state, $ja$,  weighted by the probabilities $|\psi^{}_j(x)|^2=J^2_j(x)$,
\begin{align}
\xi^2_{MS}=\sum_{j=-M}^M (ja)^2 J^{2}_j(x)\ .
    \label{xiMS}
\end{align}
In the limit $M\rightarrow\infty$ we can square the two sides of the Bessel functions'  identity (\ref{id}), sum over $j$ 
and use Eq.~(\ref{ort1}), 
yielding
\begin{align}
\xi^{}_{MS}=x a/\sqrt{2}\ .
 \label{xtilde}
\end{align}
Again, $\xi^{}_{MS}$ is proportional to $xa$, but with a different prefactor.

\section{Many-electrons occupations}    \label{MB}     

So far we ignored the spin of the electron, and considered the electronic states of a single electron. As long as the Hamiltonian does not contain the spin degrees of freedom, the results also apply when we include the spin of the electrons, and the Pauli principle implies that the single-orbital level on each lattice site 
can be occupied by 0, 1 or 2 electrons. Therefore, the total number of electrons on the lattice, $N^{}_e$, can have any value between $0$ and $2N$.
Here we consider the many-electron problem, ignoring the electron-electron interactions. In the ground state of this many-electrons system, each of the available energy levels is filled with two electrons, beginning with the lowest energy and up until the number of available electrons, $N^{}_e$, is exhausted 
at the Fermi energy.  

 In practice, we need to consider finite chains, with sites $j=-M,~-M+1,~\dots,~M$. As stated, a hard wall for $j<-M$ will modify the wave functions near that wall. At small $x$, this modification will affect only wave functions within a distance $\xi$ from the wall. Assuming $\xi\ll Ma$ (as in Fig. \ref{Fig1}),  effects near the ``edge" at $j=0$ can be neglected (for small $x$, a pertubation expansion in $t$ generates corrections to the wave functions which are of order $x$).

We now fill the  (orbital) single-electron eigenenergies ${\cal E}^{}_\ell$, with eigenfunctions $\psi^{}_\ell(j)$.   If $N^{}_e$ is even, then each such state is occupied by two electrons, and the corresponding electronic density (or ``occupation") at site $j$ is
\begin{align}
    {\rm n}(j) = 2\sum|\psi^{}_\ell(j)|^2\ ,
\end{align}
where the sum is over the $N^{}_e/2$ lowest energies,  beginning with $\ell=-M$. If $N^{}_e$ is odd, the next level contains only one electron, and one adds $|\psi^{}_{(N^{}_e+1)/2}(j)|^2$. If the single electron energies are degenerate then each $|\psi^{}_\ell(j)|^2$ will be replaced by a sum over the corresponding degenerate squared wave functions, and the sum will stop after filling $N^{}_e$ orbital states. 

In our example, the total number of orbital states is odd, $2M+1$. If each of the isolated  atoms in the lattice contains a single electron in its ``outer" level (as e.g. for alkali atoms), then there is a total of $N^{}_e=2M+1$ electrons. As the Pauli principle allows for  two electrons (of opposite spins) in each orbital,  this situation is termed  ``half-filling".


\vspace{-4mm}

\subsection{ Bloch electrons}

For the Bloch electrons,  discussed in Sec. \ref{Bloch}, each energy ${\cal E}^{}_\ell$ (except for $\ell=0$) has two orbital states, and therefore it can be occupied by four electrons. 
At half-filling, this implies occupation of all the states with $|k^{}_\ell|<\pi/(2a)$,  with energies between $-2t$ and $0$, i.e., in the 
lower half of the energy band. For large $N$, these energies form a dense quasi-continuum. At very low temperatures we can ignore the fully-occupied lower energy states, and concentrate on energy levels near the Fermi energy, which is now equal to zero. Without an external voltage, there is an equal number of electrons with positive and negative momenta. Application of a voltage causes some electrons to shift to nearby empty states with higher momenta, generating a net drift velocity that leads to an electric current.   This system  is a  ``metal". If each atom contributes 2 electrons then the whole energy band is filled with electrons, and the chain is an  ``insulator", unless there exist empty states above a small gap, to which the upper electrons can hop at finite temperatures, forming a  ``semiconductor"~\cite{book,Powell,ashcroft-mermin}.

\vspace{-4mm}

\subsection{Wannier-Stark electrons}

In the Wannier-Stark chain, the non-degenerate discrete energy levels
are ${\cal E}^{}_\ell=eaF\ell$, with $\ell=-M,~-M+1,~\ldots,~0,~\ldots,~M$. The eigenfunction $\psi^{}_\ell(j;x)$ is maximal at the site $j=\ell$.
  At half-filling, all the levels below $0$ (up to $M=-1$)  will be occupied by 2 electrons, and the  level at $0$ will contain one electron. The Fermi energy is again equal to zero.
The total occupation at the site $j$  is
\begin{align}
    {\rm n}(j;x) = 2\sum_{\ell=-M}^{-1} J^{2}_{\ell-j}(x)+ J^{2}_{j}(x)\ .
 \label{Occ}
\end{align}

\begin{figure}
\includegraphics[width=0.4\textwidth]{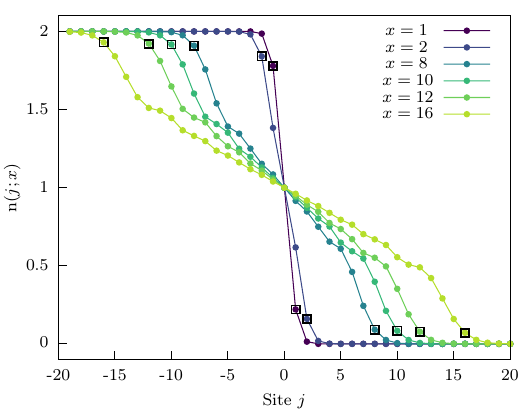}
\caption{(Color online) Occupation number profile for different values of $x$, Eq. (\ref{Occ}). The squares indicate the distance $\xi$ from the origin, see Eq. (\ref{dw}).}
\label{Occupation}
\end{figure}

The function ${\rm n}(j;x)$ is shown in Fig.~\ref{Occupation}, which also indicates the localization length $\xi$ for each value of $x$. The plots used the sum in Eq. (\ref{Occ}), truncated at $\ell=-M$. For all the curves, $\xi<Ma$, and this truncation, as well as the boundaries, do not affect the results.
As expected, the states with large negative $\ell$ are fully occupied, each by 2 electrons, and they dominate the sum in Eq. (\ref{Occ}) for large negative $j$. The eigenfunction $J^{}_{\ell-j}$ has significant contributions only from small $|\ell-j|$. Therefore, the occupation there is very close to 2. Similarly, the states with positive $\ell$ are empty, and therefore the occupation of sites with large positive $j$ is very small.
 Around the  ``edge", between these two regions, the occupation gradually decreases from $2$ to $0$, with a width that increases inversely proportional to the electric field.  In the strong electric-field limit, where $x \rightarrow 0$, $J^{}_0(x) \rightarrow 1$, $J^{}_1(x) \rightarrow 0$, and then ${\rm n}(-1;x) \rightarrow 2$, while ${\rm n}(1;x) \rightarrow 0$. This tendency of ${\rm n}(j;x)$ to be approximately a step function in the strong-field limit is indeed seen in Fig.~\ref{Occupation}. 
 
In any case, the Wannier-Stark states are all localized, and their energies have finite gaps between them. Therefore, these one-dimensional systems are always insulators. Other properties of the single electron wave-functions are discussed in Ref.~\cite{fukuyama}.



 Although Fig. \ref{Occupation} was drawn for half-filling, we would obtain exactly the same plots for other fillings, except that then the location of the ``center" of the ``edge" would move from $j=0$ to the location at which the available electrons stop filling the localized states. Thus, the chemical potential will allow us to shift the location of the boundary between ``bulk" and ``vacuum" simply by changing the number of electrons in the system.


\section{Edge States in Two Dimensions}\label{ES}

 The ``edge" states found above in the one-dimensional case are still localized. To discuss ``mobile" edge states one needs to extend the previous results  to the two-dimensional case. Accordingly, we introduce now a two-dimensional tight binding model on a cylinder, with tunneling and electric field along the horizontal $x-$axis and with only tunneling along the circumference~\cite{Khanna2022}. For a  lattice with $N\times N'$ sites, the tight-binding equations for the sites $(j,j')$ are then
\begin{align}
[{\cal E}&^{}_{\ell,\ell'}/t-2j/x]\psi^{}_{\ell,\ell'}(j,j')=\nonumber\\
&-\big[\psi^{}_{\ell,\ell'}(j-1,j')+\psi^{}_{\ell,\ell'}(j+1,j')\big]\nonumber\\
&-(t'/t)\big[\psi^{}_{\ell,\ell'}(j,j'-1)+\psi^{}_{\ell,\ell'}(j,j'+1)\big]\ ,
\label{TB2D}
\end{align}
where both the energies and the eigenfunctions depend on the dimensionless parameters $x$ and $t'/t$.  With periodic boundary conditions in the transverse direction, the solutions are
\begin{align}
\psi^{}_{\ell,\ell'}(j,j';x,t'/t) \rightarrow J^{}_{\ell-j}(x)e^{ik{}_{\ell'} a j'}/\sqrt{N'}\ ,
\label{ESWF}
\end{align}
with the  eigenvalues 
\begin{align}
{\cal E}^{}_{\ell,\ell'}(x,t'/t)/(eaF)=\ell-2[t'/(eaF)]\cos(k^{}_{\ell'} a)\ ,
\label{E2D}
\end{align}
 and with  $k^{}_{\ell'}=2\pi\ell'/N'$.  See also Exercise 8 in App. \ref{App}.  

\begin{figure}
\vspace{4mm}
\includegraphics[width=0.4\textwidth]{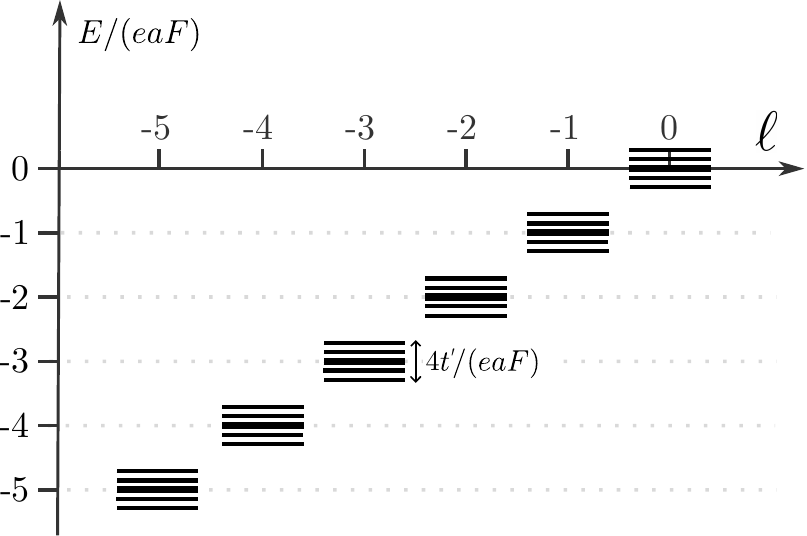}
\caption{The energy levels of the two-dimensional case, with $M=5,~N'=5$, Eq. (\ref{E2D}). The horizontal axis denotes the index $\ell$, which is also equal to the horizontal site index around which the corresponding states are localized, $j=\ell$. The thick segments represent the Wannier-Stark ladder. Each of these energies is ``dressed" by a band of $N'$ levels, representing the band of ``free" electronic states in the transverse direction. The width of such a band is indicated by the double headed vertical arrow.}
\label{Fig3}
\end{figure}

 Equation~(\ref{ESWF}) is the wave-function for a single electron in a two-dimensional  state characterized by quantum numbers $\ell$ and $\ell'$. It represents a hybrid state which is localized along the horizontal  direction labeled by $j$, but fully extended around the circumference labeled by $j'$. An example of this spectrum is shown in Fig. \ref{Fig3}. As seen in the figure, each discrete level of the Wannier-Stark ladder is now accompanied by a band of   transverse modes.   For $t'\ll eaF$ (shown in the figure), there is no overlap of energies from different ``columns". Otherwise, there may arise some degeneracies between neighboring ``longitudinal" states. In any case, these states will be gradually filled from left to right. Once all the states in a given column (i.e., for a given value of $\ell$) are filled, they cannot participate in conduction. Assuming that all the states with $\ell<0$ are filled, the Fermi energy corresponds only to states with $\ell=0$, and --  as long as the transverse states there are not fully occupied -- we can have mobility of electrons involving the transverse states with $j$ around zero. We have thus constructed a model with an ``insulating" bulk and a ``metallic" edge.

\section{Electron-electron interactions}\label{dis}

Studying the effect of electron-electron interactions requires advanced theoretical and numerical tools. Here we present only a qualitative description of these effects in a one-dimensional lattice, introducing several relevant concepts. Again, we discuss the ground state. Without the electric field, a strong repulsive onsite electron-electron interaction $U$ implies that in the ground state every site on the lattice tends to be occupied by (at most)  one electron. If this is the case, then for half-filling, all sites are expected to be singly occupied, making the system an insulator. However, virtual excited states through which an electron can tunnel to a neighboring site (``paying" an energy $U$ and ``gaining" an anergy $t$), require that electrons on neighboring sites have opposite spins (electrons with parallel spins cannot occupy the same site), and hence the system is a  ``Mott insulating antiferromagnet". Applying an electric field competes with this structure, and numerical studies show the appearance of the Mott antiferromagnet only in an intermediate region around the  edge  at $j=0$,  where the site-dependent potential is small~\cite{Khanna2022,coex}.  For zero electric field, adding also a strong repulsive electron-electron interaction between electrons on nearest-neighbor sites may prefer a state in which alternate sites are occupied by 2 or 0 electrons. This is an example of a ``charge-density wave". In the presence of an electric field, this state appears around the edge of the system ~\cite{JPC}. Combining both onsite and nearest-neighbor interactions can generate a multitude of intermediate structures around the edge~\cite{Khanna2022,coex}. More intermediate structures also arise at other levels of filling, e.g. quarter-filling.

\section{Conclusion}

This  paper introduced  the concepts of localization, confinement, bulk and edge states  through the example of the linear Wannier-Stark potential, which can be solved anlytically for very long chains. This also illustrated the use of Bessel function  in a simple context. These topics are not included in most of the standard textbooks on quantum mechanics or solid state physics, but can be included as special topics.  

\section{Author Declarations Section}

All authors contributed equally to this work.

\appendix

\section{Exercises}
\label{App}

\begin{enumerate}

\item Show that the Bloch eigenstates are orthonormal, 
\begin{align}
\langle \Psi^{}_{\ell}|\Psi^{}_{\ell'}\rangle=\sum_{j=-M}^M\psi^{}_\ell(j)^\ast\psi^{}_{\ell'}(j)=\delta^{}_{\ell,\ell'}\ .
\label{ort1}
\end{align}

\item Find the eigenfunctions and the eigenenergies of Eq. (\ref{TB}) for a ``closed system", with the ``hard walls" boundary conditions $\psi(-M-1)=\psi(M+1)=0$ (Hint: the equations for $\psi(-M)$ and $\psi(M)$ must be treated separately.) 
\item Repeat the analysis of Sec. \ref{Bloch} for an even number of sites, $N=2M$.

\item  What is the mean square localization length for the Bloch electrons?

\item   What is  the occupation number profile for the periodic tight-binding chain at half-filling?


\item  (a)  For $M\rightarrow\infty$, (a) show that ${\rm n}(j,x)$ obeys the reflection property ${\rm n}(j,x) + {\rm n}(-j,x) = 2$.
[Hint: use Eq.~(\ref{ort1})]. 
Show that this implies that ${\rm n}(0,x)=1$ for all field strengths.

(b) Use Eq. (\ref{Occ}) to prove that
     ${\rm n}(\pm 1;x) = 1\mp[J^2_0(x)+J^2_1(x)]$, as seen in Fig. \ref{Occupation}.
     
(c) Prove that
   $\sum_{j=-M}^M {\rm n}(j;x) = 2M+1 $. 
   
   \item  For even $N=2M$, it is convenient to use $j=-M,~\ldots, -1,~0,~\ldots,~M-1$. It is also convenient to shift the energy by $eaF/2$ and shift the  origin of the $x-$axis to $-1/2$ ~\cite{JPC}. Repeat the above analysis and show that the results remain similar to those for odd $N$ for $N\gg 1$.

\item (a) In the two-dimensional case discussed in Sec. \ref{dis}, what are the possible values of $k^{}_{\ell'}$ for hard wall  boundary conditions?

     (b) How are these levels filled at half-filling, when $N^{}_e=N N'$?

 \end{enumerate}
\end{document}